\documentclass[showpacs,pra,aps]{revtex4}
\usepackage[dvips]{graphics,epsfig,color}
\newcommand{\dt}[1]{\frac{d\,#1}{dt}}
\def\c#1{\cite{#1}}

\def\dfrac{\displaystyle\frac}

\newcommand{\ka}{\hbox{\ae}}

\begin{document}
\centerline{CHINESE PHYSICS, Vol. {\bf 9}, Number 2, pp. 124-130, February, 2000\\[5mm]}

\title{Inversionless gain in a three-level system driven by a strong field and collisions
\footnote{Project supported in part by the Russian Fundation for Basic Research
(Grant No.96-02-00010c),  by the National Sciences Foundation of China (Grant No.
19911121500) and the Science and Technology Foundation of Jilin Province, China
(Grant No. 980526).}}
\author{A.K.Popov and S.A.Myslivets}
\affiliation{Institute for Physics, Russian Academy of Sciences and Krasnoyarsk State
University, 660036 Krasnoyarsk, Russia. E-mail: popov@ksc.krasn.ru, fax:
(3912)438923,}
\author{Gao Jin-yue  and Zhang Han-zhuang }
\affiliation{Department of Physics, Jilin University, Changchun 130023, China}
\author{B.Wellegehausen}
\affiliation{Institut f\"ur Quantenoptik, Universit\"at Hannover, Welfengarten 1,
30167 Hannover, Germany,\\ Fax:+49-511/762-2211,
E-mail:Wellegehausen@mbox.iqo.uni-hannover.de}
\author{Received 14 March 1999}
\affiliation{}
\begin{abstract}
Inversionless gain in a  three-level system driven by a strong external field and by
collisions with a buffer gas is investigated. The mechanism of populating of the
upper laser level contributed by the collision transfer as well as by relaxation
caused by a buffer gas  is discussed in detail. Explicit formulae for analysis of
optimal conditions are derived. The mechanism developed here for the incoherent pump
could be generalized to other systems.

\pacs{4250; 3150 }
\end{abstract}
\maketitle

\section{ Introduction}

{ Light amplification without population inversion (AWI) has been studied in a number
of theoretical and experimental papers \cite{po} - \cite {mes} because of its
potential application in producing high-power lasers in the regions of
electromagnetic spectrum which are difficult to reach with traditional laser system.
Although population inversion between the laser levels is not necessary in
inversionless light amplification system, it is required to pump a fraction of
population to the upper level incoherently. Possible AWI at the transitions between
excited states of noble gases, pumped by discharge, was shown in \cite{{po},{am}}.
The incoherent pump by discharge may not always provide necessary population, and the
direct excitation by laser with broad linewidth may introduce additional atomic
coherence. Another approach to possible AWI assisted by the collisional population
transfer between fine structure levels of alkali atoms was proposed in
\cite{{cr},{fl},{wod}} and based on successful experiments \cite {{atu},{mov}} proved
that wide range manipulation by population of the levels (up to population inversion)
can be provided with this mechanism. Successful experiments on AWI in potassium
vapors based on such mechanism were reported in \cite{str}. (The theory and estimates
for sodium atoms, underlying \cite{{cr},{fl},{wod}} were presented in
\cite{{wel},{izv}} too). Hyperfine structure of D-lines in alkalies and degeneracy of
the coupled levels require multilevel model in order to consider complex of quantum
nonlinear coherence and interference effects, which are origin of AWI. However
experiments with sodium \cite{gu} and potassium \cite{{str}} atoms showed, that major
behavior and requirements for AWI (especially at detuning of the driving field from
the resonance) may be studied with the aid of simple three-level V-configuration. In
this paper, the mechanism of incoherent pump by a strong laser beam and collisional
population transfer induced by buffer gas pressure is further analyzed. In order to
fit better experimental conditions degeneracy are taken into account. We derived
simple and explicit formulae, describing process. With the aid of these expressions
we compare various alkalies and buffer gases in order to show that strong competition
between population transfer and loss of coherence may be optimize by the proper
choice of the transitions and buffer gases. The idea developed here for incoherent
pump could be generalized to other systems.
}

\section{Theoretical model and equations}

\begin{figure}[h]
\includegraphics[width=.3\textwidth]{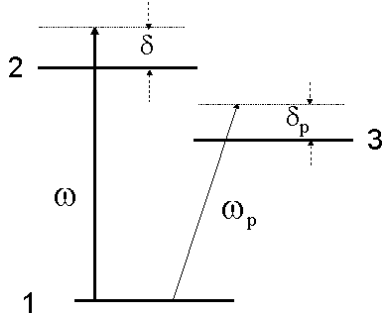}
\caption{Level scheme for
inversionless amplification
in the collision- driven three-level system. Transition ${|3\rangle -
|1\rangle}$ is probed by the week radiation at the frequency $\omega _p$.
Transition ${|2\rangle - |1\rangle}$ is coupled to the strong radiation at
the frequency $\omega $. Level ${|1\rangle}$ is ground state,
${|2\rangle}$ and ${|3\rangle}$ are fine splitting levels. In order to
manipulate  population difference at the probe transition in the wide
range from positive (absorption) via zero to negative values (gain), while
maintaining absorption at the driven transition, we consider
collisions with atoms of a buffer gas along with the
saturation effects. Near Boltzmann's population distribution between
fine-structure levels may set up due to the collisions.}
\end{figure}
{ We consider a model as shown in Fig.1 which consists of three levels
labeled $|1\rangle$, $|2\rangle$ and $|3\rangle$. A strong driving field with
frequency $\omega$ produces coherent coupling between levels $|1\rangle$ and
$|2\rangle$, while the weak probe field with
frequency $\omega_p$ is scanned around the transition between levels $%
|3\rangle$ and $|1\rangle$. We express the probe and driving fields in the
form:
\begin{eqnarray}
E_p(t)&=&\frac12[E_{p0}e^{-i\omega_pt}+c.c.]\\
E(t)&=&\frac12[E_{0}e^{-i\omega t}+c.c.] \end{eqnarray} where $E_{p0}$
and $E_0$ are the complex  amplitudes. The Hamiltonian of the
system including the interaction between the atom and the two fields can be
written as:
\begin{eqnarray}
H&=&H_a+H_b\nonumber\\ H_a&=&\hbar \omega_p a_3^+a_3+\hbar \omega a_2^+a_2\\
H_b&=&-\hbar \delta_p a_3^+a_3-\hbar \delta
a_2^+a_2-\hbar[g_pa_3^+a_1e^{-i\omega_pt}+ga_2^+a_1e^{-i\omega
t}+c.c.],\nonumber
\end{eqnarray} where $\delta_p=\omega_p-\omega_{31}$, $\delta=\omega-%
\omega_{21}$; $\omega_{31}$ and $\omega_{21}$ denote the transition
frequencies from $|3\rangle$ to $|1\rangle$ and $|2\rangle$ to $|1\rangle$,
respectively; $g=\mu_{21}E_0/2\hbar$, $g_p=\mu_{31}E_{p0}/2\hbar$ are the
Rabi frequencies of the probe and the driving fields, respectively; $%
\mu_{31} $ and $\mu_{21}$ are the dipole matrix elements of optical
transitions from $|3\rangle$ to $|1\rangle$ and $|2\rangle$ to $|1\rangle$,
respectively. In the interaction picture, the master equation for density
operator is
\begin{eqnarray}\label{r}
\dt{r}=-\frac{i}{\hbar}[\tilde{H_b},r]+\hbox{incoherent term}, \end{eqnarray}
where $r=e^{iH_at/\hbar}\rho e^{-iH_at/\hbar}$, $\tilde H_b=e^{iH_at/%
\hbar}H_b e^{-iH_at/\hbar}$. According to the equation (\ref{r}), the matrix
elements of density operator can be expressed as follows:
\begin{eqnarray}
&&\dt{r_{31}}=ig_p(r_{1}-r_{3})-(\Gamma_{31}-i\delta_p)r_{31}-igr_{32}\nonumber\\
&&\dt{r_{32}}=igr_{21}^*-ig^*r_{31}-[\Gamma_{23}-i(\delta_p-\delta)]r_{32}\nonumber\\
&&\dt{r_{21}}=ig(r_{1}-r_{2})-(\Gamma_{21}-i\delta)r_{21}-ig_pr_{32}^*
\label{eq_rn}\\
&&\dt{r_{3}}=2Im(g_p^*r_{31})-\Gamma_3r_{3}+w_{23}r_{2}\nonumber\\
&&\dt{r_{2}}=2Im(g^*r_{21})-\Gamma_2r_{2}+w_{32}r_{3}\nonumber\\
&&r_{1}+r_{2}+r_{3}=1.\label{eq_rd}
\end{eqnarray}
Here $r_{ij}=r_{ji}^*$, $r_{ii}= r_{i}$, $\Gamma_2=A_{21}+w_{23}$, $%
\Gamma_3=A_{31}+w_{32}$, where $A_{21}$ and $A_{31}$ are the decay rates of
populations in levels $|2\rangle$ and $|3\rangle$ to the ground state $%
|1\rangle$ due to the common spontaneous transition, respectively; $w_{23}$
and $w_{32}$ are the collision transfer relaxation rates of the populations
in level $|2\rangle$ to level $|3\rangle$ and in level $|3\rangle$ to level $%
|2\rangle$ due to collision; $\Gamma_{21}$, $\Gamma_{31}$, $\Gamma_{23}$ ---
are the polarization decay rates between levels $|1\rangle$ and $|2\rangle$,
levels $|3\rangle$ and $|1\rangle$, levels $|3\rangle$ and $|2\rangle$,
respectively ($\Gamma_{ij}=\Gamma_{ji}$). }

{ Absorption and refractive indices at the frequency $\omega _p$ are determined by
the complex susceptibility $\chi (\omega _p)$ which is
proportional to the off-diagonal element $r_{31}$: $\chi (\omega _p)={%
(r_{31}\mu _{13})}/({E_p/2}).$
In the liner approximation of probe field, the steady state solution of
equations (\ref{eq_rn}) for $r_{31}$
is:
\begin{eqnarray}\label{r31}
r_{31}=i\frac{g_p(r_{1}-r_{3})-gr_{32}}{\Gamma_{31}-i\delta_p}
\end{eqnarray}
The coherence $r_{32}$ itself is dependent on $r_{31}$ and $r_{21}$. By
substituting solution for $r_{32}$ to equation(\ref{r31}) one obtains:
\begin{eqnarray}
\frac{\chi}{\chi_0}=f(\delta_p)=i\Gamma_{31}\frac{[\Gamma_{23}-i(\delta_p-\delta)]
(r_{1}-r_{3})-igr_{21}^*}{[\Gamma_{23}-i(\delta_p-\delta)]
(\Gamma_{31}-i\delta_p)+|g|^2},\label{nnief}
\end{eqnarray}
where
\begin{equation}\label
{r21}r_{21}=-ig(r_1-r_2)/(\Gamma _{21}-i\delta).
\end{equation}
Eventually from (\ref{nnief}),(\ref{r21}) we obtain:
\begin{eqnarray}
f(\delta_p)=i\Gamma_{31}\frac{[\Gamma_{23}-i(\delta_p-\delta)]
(r_{1}-r_{3})-|g|^2(r_{1}-r_{2})/(\Gamma_{21}+i\delta)}{[\Gamma_{23}-i(\delta_p-\delta)]
(\Gamma_{31}-i\delta_p)+|g|^2},\label{f}
\end{eqnarray}
\begin{eqnarray*}
{\alpha_p}/{\alpha_{p0}}=Im\,f(\delta_p),\>
({n_p-1})/({n_{p\,max}-1})=Re\,f(\delta_p)
\end{eqnarray*}
}

\section{ Mechanism of power and collision driven $AWI$}

{Optimum conditions for $AWI$ in three-level $V$ configuration, assuming possible
manipulating by populations of the coupled levels with incoherent pump were analyzed
in \cite{po}. Classification of the effects of a strong field on the spectral line
shape at an adjacent transition for various $V$, $\Lambda$ and $H$ configurations of
coupled transitions was given in \cite{{po},{sok}} (see also
\cite{rus}(e),\cite{he}): dependencies of $r_{i,j}$ and of the denominator in
(\ref{nnief}) on $g$ refer to saturation of populations and energy-level splitting
effects, correspondingly; $r_{21}$-- represents nonlinear interference effects
($NIEF$). As it was shown in \cite{sok},\cite{rus}(e) NIEF bring about change of a
line shape but not an integral intensity:
\begin{equation}
\int d\delta_pf(\delta_p,|E|^{2}) = \pi\Gamma_{31}(r_{1}- r_{3}),
\end{equation}
\noindent which depends only on saturation effects. }

{Indeed, NIEF represent the origin of sign-changing spectral line shapes and AWI at
the probe transition. One can regard (\ref{nnief}) as the difference between acts of
pure emission (associated with the $r_{3}$, assuming $r_{1}= r_{2}= 0$), and pure
absorption (the rest terms). Both of them are positive but depend in a different way
on detunings, because of NIEF. This was emphasized in \cite{sok} (see also
\cite{rus}(e),\cite {{po},{he}}). }

{Thus, in the schematic under consideration $AWI$ originates from the coherence at
the transition $|3\rangle-|2\rangle$, induced by the strong field, coupled to the
auxiliary level $|2\rangle$ in combination with the
probe field, which consequently gives rise to factor $r_{21}$ in (\ref{nnief}%
). The larger is maximum value of $r_{21}$ compared to $r_{1}-r_{3}$ the
more pronounced is the effect of $AWI$.
\noindent At $\delta = 0,$ maximum of absorption (gain) corresponds to
$\delta_p = 0.$
\begin{eqnarray}
f(0)=\frac{(r_{1}-r_{3})-(r_{1}-r_{2})S}{1+S},
\label{fo}
\end{eqnarray}
where $S=|g|^2/\Gamma_{21}\Gamma_{32}$. 
\noindent Therefore, even at $(r_{1} - r_{3})>0$, $(r_{1} - r_{2})>0,$
negative absorption (gain) occurs if
\begin{eqnarray}
(r_{1} - r_{2})S > r_{1} - r_{3}.
\end{eqnarray}
\noindent The less is coherence decay rate $\Gamma_{32}$ at the two-photon
transition $|3\rangle-|2\rangle$, compared to that at the coupled one-photon
transitions, the more favorable are the conditions for AWI. At $S > 1$ large
splitting of the level $|1\rangle$ on two quasi levels significantly reduces
interference and, therefore, the magnitude of AWI at the center of the
transition $|3\rangle-|1\rangle$. Optimal value of strength of driving field
for $AWI$ in the probe line center was analyzed in \cite{po}. }

{Below we shall consider opportunities and specific experimental schematic to control
shape of absorption and refraction indexes without external incoherent pump.
Necessary for $AWI$ distribution of populations is ensured by collisions. }

{ Consider alkali atoms, immersed into buffer gas. Strong field couples $P_{3/2}$ and
ground $S$ levels. Fast collision exchange ensures population transfer from the
$P_{3/2}$ to the lower $P_{1/2}$ level. In order to understand main mechanism giving
rise to $AWI$, suppose for simplicity, that pressure of the buffer gas is so strong,
that Boltzmann population distribution between the fine structure levels is
established. This brings about the means to control population difference at the probe
transition in a wide range by increasing intensity of the strong field. Even
population inversion at the $P_{1/2} -S$ transition can be provided due to the
saturation effect at $P_{3/2} - S$ transition (similar to that in Ruby laser). Such a
feasibility to produce population inversion at the transition $P_{1/2} -S$ has been
demonstrated experimentally for potassium \cite{mov} and sodium \cite{atu} vapors
admixed to helium at about atmospheric pressure. Possible $AWI$, the required driving
field strength and buffer gas
pressure can be estimated as follows. One can find from the equation system (%
\ref{eq_rd}):
\begin{equation}
\label{r'}
r_{2} ={\frac{\Gamma_3}{2\Gamma_3+w_{23}}}\cdot{\frac{\hbox{\ae}^{\prime}}{1+%
\hbox{\ae}^{\prime}}},\> r_{3} =\frac{w_{23}}{\Gamma_3}r_{2},\> r_{1} =(1+%
\frac{\Gamma_3\hbox{\ae}^{\prime}}{2\Gamma_3+w_{23}})\frac{1}{1+\hbox{\ae}%
^{\prime}};
\end{equation}
\begin{equation}
\label{dr}r_{1}-r_{2} =\frac{1}{1+\hbox{\ae}^{\prime}},\> r_{1}-r_{3} =(1+%
\frac{\Gamma_3-w_{23}}{2\Gamma_3+w_{23}}\hbox{\ae}^{\prime})\frac{1}{1+%
\hbox{\ae}^{\prime}}.
\end{equation}
Here
\begin{eqnarray}
\ka'=\frac{2+(w_{23}/\Gamma_3)}{[1-(w_{32}w_{23}/\Gamma_3\Gamma_{2})}\cdot\ka,
\label{ae'}
\end{eqnarray}
\begin{eqnarray}
\ka=2\Gamma|g|^2/\Gamma_2(\Gamma^2+\delta^2)
\label{ae}
\end{eqnarray}
}

{ Consider high pressure limit, assuming that $A_{31} \simeq A_{21}$, and the buffer
gas pressure is so high, that $(w_{23}-w_{32})
>>A_{31},A_{21} $. Then taking into account that in the given conditions $%
w_{32} = w_{23}\exp (-\Delta E/k_{B}T)$, where $\Delta E = E_{2} - E_{3}$ is
the fine splitting energy, $k_{B}$ and $T$ are Boltzmann's constant and
temperature, we obtain:
\begin{eqnarray}
{r_{1}-r_{3}}=\left[1-\ka'\frac{1-\exp(-\Delta E/k_{B} T)}
{1+2\exp(-\Delta E/k_{B} T)}\right]\frac1{1+\ka'};\label{8} \\
\ka'=\frac{1+2\exp(-\Delta E/k_{B} T)}{1+ \exp(-\Delta E/k_{B} T)}
\cdot\frac{2 |g|^2\Gamma_{21}/A_{21}}{\Gamma_{21}^2+\delta^2}.
\label{9}
\end{eqnarray}
For sodium $\Delta E = 17.2\, \mbox{cm$^{-1}$ and at }T = 550\,K$ the
estimates give
$$
\Delta E/k_{B}T = 4.3\cdot 10^{-2},\hspace{2ex} \hbox{\ae}^{\prime}\simeq 3
|g|^{2}/\Gamma\Gamma_{2} \simeq 9\lambda^{3} I/64\pi^{3} \epsilon_{0}\hbar
c\Gamma_{21},
$$
\begin{equation}
\label{10}r_{1}-r_{3}\simeq \frac1{1+\hbox{\ae}^{\prime}}[1-1.3\cdot10^{-2}%
\hbox{\ae}^{\prime}]
\end{equation}
Here $\lambda$ and I are strong field wavelength and energy flux density, $%
\epsilon _{0}$ is primitivity of free space. From the equations (\ref{fo}),(%
\ref{8}),(\ref{9}) one can see that the potentially attainable AWI grows
with the increase of $\Delta E$ (for example, in $K$ and $Rb)$. But on the
other hand with the increase of fine splitting population transferring
collision crossection decreases. That must be compensated with increase of
buffer gas pressure. Later brings loss of coherence. These dependencies will
be investigated numerically in the next section. }

{ Inelastic collision crossection of sodium and helium for the
transition $3P_{3/2}- 3P_{1/2}$ is $\sigma _{23} \simeq 4\cdot 10^{-15}$ cm$%
^{2}$ \cite{11}. For T= 550K and atmospheric pressure of helium we estimate $%
w_{23}= N_{He}\bar{v}\sigma _{23}\simeq 7.5\cdot 10^{9}$ sec$^{-1}$. Since $%
A_{31}\simeq A_{21}\simeq 6.2\cdot 10^{7}$ sec$^{-1}$, requirements for the
approximations (\ref{8}),(\ref{9}) are met. Taking the data of ref.\cite{12}
for the collision broadening of sodium $D$ lines by helium, we estimate
collision halfwidth as $\Gamma_{21} \simeq 5\cdot 10^{10}$ sec$^{-1}$, which
exceeds measured Doppler halfwidth of this transition $\Delta \omega _{D}/2
=4.7\cdot 10^{9}$ sec$^{-1}$ $(\Delta \nu_{D}/2 =0.75$ GHz). So we can
neglect inhomogeneous broadening of the transition. }

{ For the conditions under consideration we estimate $\hbox{\ae}%
^{\prime}\simeq 5\cdot 10^{-9}\cdot I$ (where I is in W/cm$^{2}$), $%
|g|^{2}/\Gamma_{21}\Gamma_{32}\simeq |g|^{2}/\Gamma_{31}\Gamma_{32} \simeq
\hbox{\ae}^{\prime}\Gamma_{2}/3\Gamma_{32}$. For the laser power $0.1$ W
focused to the spot $A = 10^{-5}\,\mbox{cm$^{2}$}$ (confocal parameter $b
\simeq 1$ cm) we obtain $|g| \simeq 3.6$ GHz, $\hbox{\ae}^{\prime}\simeq
5\cdot 10^{2}$, $|g|^{2}/\Gamma_{21}\Gamma_{32}\simeq 0.1$. These magnitudes
are about optimal ones and correspond to the estimated values, required to
vary population difference at the probe transition $r_{1}- r_{3}$ around
zero. Above presented estimates for the intensity $(1-10)\,\mbox{kW/cm$^{2}$}
$, required to achieve appreciated change of the line shape under the
conditions considered, compares well with the experimental data obtained for
the significant change of the ratio of population differences at the coupled
transitions at the similar conditions \cite{atu}. }

{ The magnitude of AWI at $r_{1}-r_{3}= 0$ is estimated as $%
\alpha_{p}(0)/ \alpha_{p0} \simeq \Gamma_{2}/3\Gamma_{32}$. Assuming $%
\Gamma_{32}\simeq w_{23}$, it may yield about $0.3\%$ of the absorption in
the absence of the strong field. It is seen that this quantity is very
sensitive to the decay rate of the coherence at the Raman-like transition $%
|3\rangle-|2\rangle$. 
}

\section{  Numerical analysis of power and collision induced
AWI at the transition between degenerate levels of alkalies}

{ We shall consider such frequency detunings and collision broadening of the
transitions, that hyperfine splitting of the levels can be neglected. However,
account for degenerating of the levels may occur important for the quantitative
analysis of the optimum conditions for $AWI$. }

{ Assume that both driving and probe fields are linear and in the same way polarized.
Then we assume that collisions can transfer only populations but not coherence. By
that we can consider equations (\ref{eq_rn}) for off-diagonal elements of density
matrix as referred to the transitions between the same Zeeman sublevels of the upper
and lower energy levels, which are not coupled. Unlike that equations for populations
of the sublevels are coupled by the collisions. We assume that collisions are so
strong, that populations of all sublevels within a same level are equal. Consequently
we shall consider populations of any sublevel $r_{iM}$ and of the level $R_i$ to be
related as $r_{iM} = R_i/g_i$, where $g_i$ is degenerating factor of the level.
Equations for the populations take the form:
\begin{eqnarray}
R_2=({g_2\over g_1}R_1-R_2)\ka +R_3 {w_{32}\over \Gamma_2};\>
R_3= R_2{w_{23}\over\Gamma_3};
R_1+R_2+R_3=1,
\label{R}
\end{eqnarray}
where $\hbox{\ae}$ is given by (\ref{ae}). Solution of (\ref{R}) is:
\begin{eqnarray}
&R_2=\dfrac{g_2/ g_1}{1+(g_2/ g_1)[1+(w_{23}/\Gamma_3)]}
\cdot\dfrac{\ka'}{1+\ka'};\>  R_3= R_2{w_{23}\over\Gamma_3};&\nonumber\\
&R_1=\left[1+\dfrac{\ka'}{1+(g_2/ g_1)[1+(w_{23}/\Gamma_3)]}\right]
\cdot\dfrac{1}{1+\ka'},&\label{R'}\\
&\dfrac{R_1}{g_1}-\dfrac{R_2}{g_2}=\dfrac{1}{g_1}\cdot{1\over 1+\ka'},&
\label{R12}\\
&\dfrac{R_1}{g_1}-\dfrac{R_3}{g_3}=\dfrac{1}{g_1}\left\{1+
\dfrac{[1-(g_2w_{23}/g_3\Gamma_3)]\ka'}{1+(g_2/ g_1)[1+(w_{23}/\Gamma_3)]}
\right\}\cdot\dfrac{1}{1+\ka'}.&
\label{R13}
\end{eqnarray}
It is seen from (\ref{R13}) that population inversion at the probe
transition is impossible until buffer gas number density $N_b$ meets
requirement:
\begin{equation}
\label{inv}N_b\langle v \rangle(g_2\sigma_{23}-g_3\sigma_{32})<g_3A_{31}
\end{equation}
Here $\langle v \rangle$ and $\sigma_{ij}$ are averaged relative collision
velocity and fine structure population transfer crossections with the buffer
atom. }

 For numerical analysis we use data presented in Table 1 (with $He$ as buffer gas).
\begin{table}[h]\caption{Collision cross-sections}
\begin{tabular}{ccccc}
Atom&$\sigma_{P_{3/2}-P_{1/2}}$(\AA$^2$)&$\sigma_{P_{1/2}-P_{3/2}}$(\AA$^2$)&$\sigma_{S_{1/2}-P_{3
/2}}$(\AA$^2$)&$\sigma_{S_{1/2}-P_{1/2}}$(\AA$^2$)\\ \hline
Na&41.1\c{ru}&77\c{ru}&159\c{lew}&137\c{lew}\\ 
K&52.8\c{ru}&84\c{ru}&133\c{lew}&100\c{lew}\\ 
Rb&0.12\c{ru}&0.1\c{ru}&145\c{lew}&145\c{lew}\\ \hline
\end{tabular}
\end{table}
\begin{figure}[h]
\includegraphics[width=.75\textwidth]{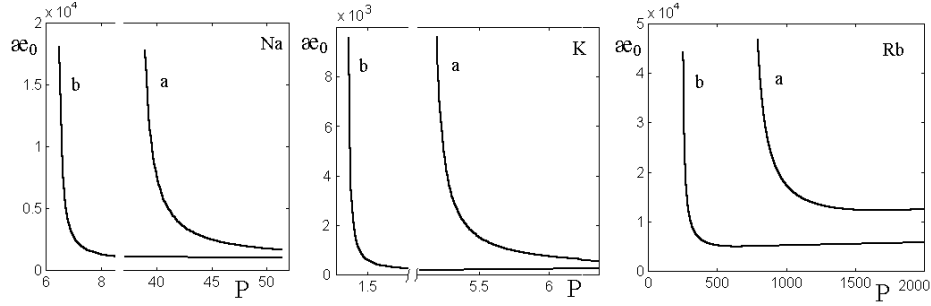}
\caption{Threshold value of the saturation parameter $\hbox{\ae}_0$, required for
population inversion to be achieved, vs buffer gas ({\it He}) pressure (a). Threshold
value of the saturation parameter $\hbox{\ae}_0$, required for inversion of sign of
$\alpha_p$ ($AWI$) (AWI) to be achieved, vs {\it He} pressure (b). Both dependencies
are drown for  atoms at $v=0$ and at $\delta=\delta_p=0$. $\hbox{\ae}_0$
 is resonant value for saturation parameter at collisionless limit
($\hbox{\ae}_0=4|g|^2/ A_{21}^2$). }
\end{figure}
 Fig.2 (curves a) shows dependence of threshold value of
the saturation parameter $\hbox{\ae}_0$, required for population inversion
to be achieved, on the buffer gas pressure above this limit (equation ${%
(R_1/g_1)-(R_3/ g_3)}=0$). Curves b show dependence of threshold value of
the saturation parameter $\hbox{\ae}_0$, required in order to achieve
inversion of sign of $\alpha_p$ ($AWI$), on the buffer gas pressure
(equation $[(R_1/ g_1)-(R_3/ g_3)]- [(R_1/ g_1)-(R_2/g_2)]S=0$). Both
dependencies are drown for  atoms at $v=0$ and at $%
\delta=\delta_p=0$. $\hbox{\ae}_0$ is resonant value for saturation
parameter at collisionless limit ($\hbox{\ae}_0=4|g|^2/ A_{21}^2$). In the
range between curves (a) and (b)  change of sign of $\alpha_p$ occurs
without changing of sign of population difference $(R_1/ g_1)-(R_3/ g_3)$.
Minimum value of the saturation parameter $\hbox{\ae}_0$, required in order
to achieve AWI is $\hbox{\ae}_0=92$ ($P_b\simeq 3.1$Torr), $\hbox{\ae}_0=848$
($P_b\simeq 12.4$Torr) and $\hbox{\ae}_0=4971$ ($P_b\simeq 640.5$Torr) for $%
K $, $Na$ and $Rb$ respectively. Such values of the saturation parameter
lead to considerable change of level populations which becomes $R_1=0.28$%
, $R_2=0.5$, $R_3=0.22$ for $K$, $R_1=0.264$, $R_2=0.51$, $R_3=0.226$ for $%
Na $ and $R_1=0.324$, $R_2=0.462$, $R_3=0.214$ for $Rb$.

{ One can manipulate  the shape of the absorption/gain spectral line, position and
bandwidth of the absorption and gain frequency-intervals. Analysis is given in
ref.~\cite{po}. Line shape is quite sensitive to the intensity and frequency
detunings of the strong field from the resonance. Bandwidth of the gain grows and the
maximum value decreases with the increase of the intensity of the strong field above
a certain magnitude. Saturation of the population difference at the strong field
transition and splitting of the common energy level oppose AWI effect, so that it can
be optimize by the proper choice of intensity and detunings of the strong field. In
the case under consideration inelastic collision frequency is important parameter to
be optimized as well.}

{ Fig.3 demonstrate dependencies of absorption/gain of probe field at frequency
$\omega_p$ vs scaled detuning $\delta_p/\Gamma_{31}$ ($\delta=0$) under optimal
values of the buffer gas pressure and of the driving field intensity. Maximum values
of the AWI index are: $\alpha_p=0.002$  for $Na $ ($\hbox{\ae}_0=8700,\ P=170$ Torr);
$\alpha_p=0.01$ for $K$ ($\hbox{\ae}_0=370,\ P=16$ Torr) and
$\alpha_p=1.3\cdot10^{-4} $ for $Rb$ ($\hbox{\ae}_0=12300,\ P=1800$ Torr).
Corresponding population differences  $(R_1/ g_1)-(R_3/ g_3)$ are $2\cdot 10^{-5}$,
$3\cdot 10^{-4}$ and $1.2\cdot 10^{-4}$. }
\begin{figure}[!h]
\includegraphics[width=.75\textwidth]{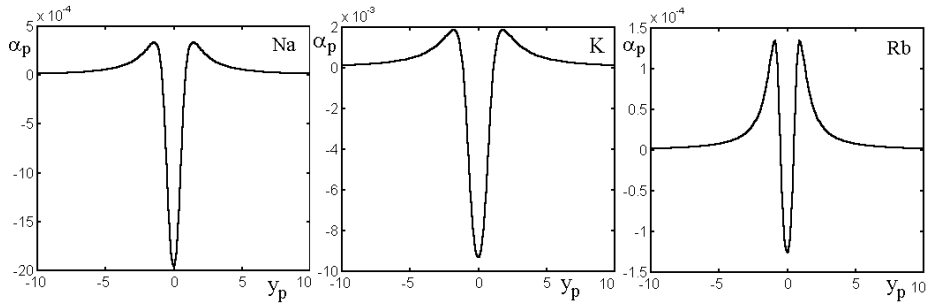}
\caption{Line shape of absorption(positive)/gain(negative) index
$\alpha_p(\delta_p$)
for the probe field at the $D_{1}$ transition in the presence of a
strong laser field, resonant to the $D_{2}$ transition. Abscisse:
detuning of the probe laser in
units of collision-broadened halfwidth $\delta_p/\Gamma_{31}$.
Ordinate: index scaled to the absorption maximum in
absence of the driving field. Strong field is tuned to exact resonance.
Curves correspond to near zero population difference at the probe
transition, so that line shapes are almost completely determined by the
NIEF. Intensity of the strong field is such that Rabi frequency $g =
2.9\cdot 10^9\hbox{sec}^{-1}$ ($Na$), $g = 0.37\cdot 10^9\hbox{sec}^{-1}$ ($%
K $)and $g = 2.2\cdot 10^9\hbox{sec}^{-1}$ ($Rb$) so that population
difference at the probe transition $|1\rangle-|3\rangle$ is still positive. }
\end{figure}

{ Taking the data of ref.\cite{lew} for the collision broadening of $D$ lines by
helium, we estimate collision halfwidth as $\Gamma_{21} \simeq 8.8\cdot 10^{9}$
sec$^{-1}$, $\Gamma_{21} \simeq 0.73\cdot 10^{9}$ sec$^{-1}$ and $\Gamma_{21} \simeq
79.8\cdot 10^{9}$ sec$^{-1}$ for $Na$, $K$ and $Rb$ respectively, which exceeds
measured Doppler halfwidth of this transition
for $Na$ ($\Delta \omega _{D}/2 =5.6\cdot 10^{9}$ sec$^{-1}$) and $Rb$ ($%
\Delta \omega _{D}/2 =2.15\cdot 10^{9}$ sec$^{-1}$). So in this case we can
neglect inhomogeneous broadening of the transition. But in the case of $K$ $%
\Gamma_{21}\ll\Delta \omega _{D}/2=3.1\cdot 10^{9}$ and it is necessary to
perform averaging over velocities of atoms.

\begin{figure}[h]
\includegraphics[width=.3\textwidth]{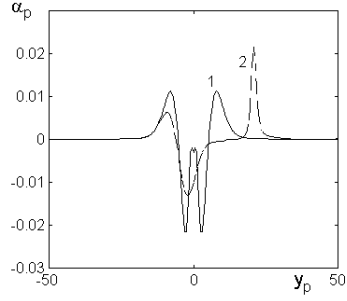}
\caption{Velocity-averaged absorption index (for $K$) at inhomogeneously broadened
$|1\rangle-|3\rangle$ transition in the presence of driving field at
$|1\rangle-|2\rangle$ transition (1 --- $\hbox{\ae}_0=3400$, $\delta=0$, 2 ---
$\hbox{\ae}_0=6\cdot10^4$, $\delta=3\cdot \Delta \omega_{D}/2$). }
\end{figure}
{ Fig.4 shows velocity-averaged absorption indices (for
$K$) at inhomogeneously broadened $|1\rangle-|3\rangle$ transition in the presence of
strong field at $|1\rangle-|2\rangle$ transition (1 --- $\hbox{\ae}_0=3400$,
$\delta=0$, 2 --- $\hbox{\ae}_0=6\cdot10^4$, $\delta=3\cdot \Delta \omega _{D}/2$).

The shape of the refraction index (dispersion) is described by the $Re f(\delta_p,
|E|^{2})$. In the scope of the discussed experiment one can manipulate by the shape
of both absorption (gain) and refraction indexes so, that maxima and minima of the
refraction would fall to the spectrum range of the near vanishing absorption. \\}

{ {\bf In summary}, this work considers a model of interference and collision driven $V$ --
type three degenerate-level system which, provides gain without population inversion
and resonance-enhanced refraction at vanishing absorption. Explicit formulas for
analyzing of the optimal conditions are presented. Comparative analysis of use of
various alkalies relative to potential experiment is qualitatively discussed in
details. Coherence destroying collision decrease the AWI and ERWA effects compared to
that in atomic beams. However the decrease may be comparable to that due to the
Doppler broadening in metal vapors. The advantages are simplicity of the experiment,
feasibilities to manipulate the population differences at the coupled transitions and
to avoid some side effects. These make the
experiment conformable to the simple common accepted theoretical model.}

\end{document}